\documentstyle[epsfig]{aipproc2}
\begin{document}
\title{The Planck Surveyor mission: astrophysical prospects }

\author{Gianfranco De Zotti$^{*}$, Luigi Toffolatti$^{*,\dagger}$,
Francisco Arg\"ueso$^{\diamondsuit}$,
Rodney D. Davies$^{\ddagger}$, Pasquale Mazzotta$^{\star \star}$, R. Bruce
Partridge$^{\odot}$, George F. Smoot$^{\star}$, and Nicola
Vittorio$^{\star \star}$}
\address{$^*$Osservatorio Astronomico di Padova, Vicolo
dell'Osservatorio 5, I-35122 Padova, Italy \\
$^{\dagger}$Dep. de F\'\i{sica}, Universidad de Oviedo, c.le Calvo
Sotelo s/n,
E-33007 Oviedo, Spain \\
$^{\diamondsuit}$Dep. de Matem\'aticas, Universidad de Oviedo, c.le
Calvo Sotelo s/n, E-33007 Oviedo, Spain \\
$^{\ddagger}$Nuffield Radio Astron. Lab., Univ. of Manchester,
Jodrell Bank, Macclesfield Cheshire SK11 9DL, UK \\
$^{\star}$LBNL, SSL, Physics Department, University of California,
Berkeley,
CA 94720, USA \\
$^{\star \star}$ Universit\`a di Roma ``Tor Vergata'', Via della Ricerca
Scientifica 1, I-00133 Roma, Italy \\
$^{\odot}$Haverford College, Haverford, PA 19041-1392, USA}

%\lefthead{LEFT head}
%\rig436thead{RIGHT head}
\maketitle

\begin{abstract}

Although the Planck Surveyor mission is optimized to map the cosmic
microwave
background anisotropies, it will also provide extremely valuable information
on astrophysical phenomena. We review our present understanding of
Galactic and extragalactic foregrounds relevant to the mission and
discuss on one side, Planck's impact on the study of their properties and,
on the other side,
to what extent foreground contamination may affect
Planck's ability to accurately determine cosmological parameters. Planck's
multifrequency surveys will be unique in their coverage of large areas
of the sky (actually, of the full sky); this will
extend by two or more orders of magnitude
the flux density interval over which mm/sub-mm counts of extragalactic
sources can be determined by instruments already available (like SCUBA) or
planned for the next decade (like the LSA-MMA or the space mission FIRST),
which go much deeper but over very limited areas. Planck will
thus provide essential complementary information on the epoch-dependent
luminosity functions. Bright radio sources will be studied over a
poorly explored frequency range where spectral signatures, essential to
understand the physical processes that are going on, show up. The
Sunyaev-Zeldovich effect, with its extremely rich information content,
 will be observed in the direction of a large number of rich clusters of
Galaxies. Thanks again to its all sky coverage,
Planck will provide unique information
on the structure and on the emission properties of the interstellar
medium in the Galaxy. At the same time, the foregrounds are unlikely to
substantially limit Planck's ability to measure the cosmological signals.
Even measurements of polarization of the primordial Cosmic Microwave
background fluctuations appear to be feasible.

\end{abstract}

\def\lsim{\, \lower2truept\hbox{${<
\atop\hbox{\raise4truept\hbox{$\sim$}}}$}\,}
\def\gsim{\, \lower2truept\hbox{${>
\atop\hbox{\raise4truept\hbox{$\sim$}}}$}\,}

\section*{Introduction}

The basic scientific goal of the Planck Surveyor mission is to measure
the cosmic microwave background (CMB) anisotropies at all angular scales
larger than $5'$, with an accuracy set by astrophysical limits. The
mission will produce all-sky maps in 9 frequency bands centered at 30,
44, 70, 100, 143, 217, 353, 545, and 857 GHz, with high sensitivity
to intensity fluctuations as
well as to polarization fluctuations, accurate calibration (through the
modulation of the $\sim 3\,$mK CMB dipole by the orbital motion of the
Earth), and a careful minimization of systematic errors.
The wide frequency coverage is achieved with two types of detectors:
tuned radio receivers at low frequencies (30 to 100 GHz: Low Frequency
Instrument or LFI), and bolometers at high frequencies (100 to 857 GHz:
High Frequency Instrument or HFI); the 100 GHz band is common to both
instruments.

The selected frequency range corresponds to a ``cosmological window'',
where foreground emissions are minimum, while the CMB intensity
peaks. At high galactic latitudes CMB
intensity fluctuations
stand out well above any other astrophysical signal; only in a tiny fraction
of pixels in Planck's high Galactic latitude maps, will the cosmological
signal
be severely contaminated. There are also good prospects for
CMB polarization measurements, especially in regions around the ecliptic
polar caps, where sensitivities significantly above average are reached.

On the other hand,
in spite of the small telescope size, resulting in a low sensitivity
to astrophysical sources compared to ground based instruments such as
the Effelsberg and Green Bank $100\,$m telescopes, the VLA, SCUBA on
the JCMT, and the mm and sub-mm arrays planned for the next decade, as
well as to space-based missions such as FIRST, the Planck surveys will
provide unique information on a broad variety of astrophysical phenomena,
including synchrotron, free-free and dust emission in our own Galaxy,
emissions from several classes of Galactic and extragalactic sources,
the Sunyaev-Zeldovich effect in clusters of Galaxies.

\section*{Foregrounds at the Planck Surveyor frequencies}

\subsection*{The Galaxy}

There are three diffuse Galactic emissions --
synchrotron, free-free, and dust --
which provide a significant confusing foreground to CMB anisotropy
observations. A sketch of the contribution of these components
to rms fluctuations at $|b|>30^\circ$ is shown as a function of frequency by
\cite{Kogut}; at high galactic latitudes, the anisotropy in
Galactic emission was found to reach a minimum between 50 and 90 GHz.

Particularly
at low Galactic latitudes the synchrotron and free-free emissions
are recognized as individual radio sources mostly
associated with recent star formation: supernova remnants and HII regions,
respectively.

\subsubsection*{Synchrotron emission}

Synchrotron emission results from cosmic-ray electrons
accelerated in magnetic fields,
and thus depends on the energy spectrum of the electrons and
the intensity of the magnetic field \cite{rybickilightman,longair}.
A recent, careful discussion of the synchrotron emission from the Galaxy
has been presented by \cite{davieswilkinson}.

If the electrons' direction of motion is random with respect to the
magnetic field, and the electrons' energy spectrum can be approximated
as a power law: $ dN /dE = N_0 E^{-p}$, then the luminosity is given by
\begin{equation}
I(\nu) \propto L N_0 B^{(p+1)/2}_{\rm eff} \nu^{-(p-1)/2} a(p) ,
\end{equation}
where $a(p)$ is a weak function of the electron energy
spectrum\cite{longair},
$L$ is the length along the line of sight through the emitting volume,
$B$ is the magnetic field strength, and $\nu$ is the frequency.

At GeV energies,
the local energy spectrum of the electrons has been measured to be,
for the energy intervals describing the peak of radio synchrotron emission,
a power law to good approximation.
The index of the power law appears to increase from about
2.7 to 3.3 over this energy range\cite{Webber,Nishimura}.
Such an increase of the electron energy spectrum slope is expected,
as the energy loss mechanisms for electrons increases
rapidly (as the square) with energy.

The synchrotron emission at frequency $\nu$ is dominated by cosmic ray
electrons of energy $E \approx 3({\nu/{\rm GHz}})^{1/2}\;{\rm GeV}$.
The observed steepening of the electrons' spectrum at GeV
energies is used to model the radio emission spectrum at GHz frequencies
\cite{Banday}.

Radio surveys at frequencies less than 2 GHz are dominated by synchrotron
emission. The well-known survey by
Haslam et al.\cite{Haslam} at 408 MHz is the only all-sky map
available.
Large-area surveys with careful attention to baselines and calibration
have also been made at 1420 MHz \cite{Reich} and most recently at
2326 MHz \cite{Jonas}.
All these investigations have been made with FWHP beamwidths of less
than 1$^\circ$.
Of particular interest for the Planck mission are surveys
of significant areas of the sky at higher frequencies, say 5-15 GHz.
Only limited data close to the Galactic plane or in selected areas
at higher latitudes with restricted angular resolution are available.
Surveys of about one steradian of the sky have been made at 5 GHz
by \cite{Melhuish} at Jodrell Bank and
at 10, 15 and 33 GHz by \cite{Gutierrez} with the Tenerife experiments.

Large features with a synchrotron spectrum extend far from the Galactic
plane.
The most prominent of these are the spurs and loops which
describe small circles on the sky with diameters in the range 60$^\circ$ to
120$^\circ$ \cite{Berkhuijsen}.
Because of their association with HI and in some cases with X-ray emission,
they are believed to be low surface brightness counterparts
of the brighter supernova remnants seen in lower latitude surveys
such as that by \cite{Duncan95} at 2.4 GHz.
Other more diffuse structure at higher latitudes may be even older remnants.

The spectral index of Galactic synchrotron emission can be readily
determined
at frequencies less than 1 GHz where the observational baselevel uncertainty
is
much less than the total Galactic emission. Lawson et al. \cite{Lawson}
used data covering the range 38 to 1420 MHz to determine
the spectral index variation over the northern sky.
Clear variations in spectral index of at least 0.3 were found.
A steepening in the spectral index at higher frequencies is observed
in the brighter features such as the loops and some SNRs.
Up to 1420 MHz no such steepening was found in the regions of weaker
emission.
At the higher Galactic latitudes where no reliable zero level is available
at 1420 MHz, an estimate can be made of the spectral index of local features
by using the T-T technique.
The de-striped 408 and 1420 MHz maps gave temperature spectral indices
of 2.8 to 3.2 in the northern galactic pole regions \cite{Davies}.

At frequencies higher than 2-3 GHz no reliable zero levels have been
measured
for large area surveys.
The relevant observational material comes from beamswitching
or interferometric data.
Accordingly the spectrum of individual features is estimated.
On scales of a few degrees the temperature spectral index between 5 GHz
\cite{Melhuish} and 408 MHz was $\sim 3.0$.
Similarly the 10, 15 and 33 GHz beam switching data \cite{Hancock}
also indicated a spectral index of $\sim 3.0$ at 10 GHz.
These results show that at higher Galactic latitudes synchrotron emission
dominates up to 5 GHz and most likely to 10 GHz.

Platania et al. \cite{Platania} used high frequency data from a number
of sky locations to investigate the frequency dependence of the
synchroton emission spectral index and also find evidence for a steepening.
Based on the local cosmic ray electron energy spectrum,
the synchrotron spectrum should steepen with frequency
to about $3.1$ at these higher microwave frequencies.

\subsubsection*{Free-free emission}

The Galactic free-free emission (see \cite{Smoot,Bartlett} for reviews)
is the thermal bremsstrahlung from hot ($\gsim 10^4\,$K)
electrons produced in the interstellar gas by the UV radiation field.
This emission is not easily indentified at radio frequencies,
except near the Galactic plane.
At higher latitudes it must be separated from synchrotron emission
by virtue of their different spectral indices.
At frequencies less than about 1 GHz, synchrotron emission dominates
at intermediate and high latitudes.
At higher frequencies where free-free emission might be expected to exceed
the synchrotron component,
the signals are weak and survey zero levels are indeterminate.

%Most of the information currently available at intermediate and high
latitudes
%comes from H$\alpha$ surveys.
Diffuse Galactic H$\alpha$ is thought to be a good tracer
of diffuse free-free emission
since both are emitted by the same ionized medium and
both have intensities proportional to emission measure
($\propto EM \equiv \int N_e^2 dl$),
the line of sight integral of the free electron density squared.
Unfortunately, no full sky map in H$\alpha$ is presently available.
Observations of limited areas of the sky have
been carried out with a Fabry-Perot spectrometer
\cite{Reynolds90,Reynolds,Marcelin} or with narrow band filters
\cite{Gaustad,Simonetti}. Note that H$\alpha$ maps are contaminated
by a strong geocoronal emission in the same line and by the OH night-sky
line
at $6568.78\,${\AA}. High resolution spectroscopy is required to separate
the Galactic H$\alpha$ emission from the contaminating lines, taking
advantage
of the separation produced by the Doppler shift due to the Earth's motion;
the separation, however, is negligible near the ecliptic poles.

The major H$\alpha$ structures form the well-known Local (Gould Belt)
System which extends 30$^\circ$-40$^\circ$ from the plane at positive $b$
in the Galactic centre and at negative latitude in the anticentre.
The HI and dust in the Local System may be traced to 50$^\circ$
from the Galactic plane.
Other H$\alpha$ features are also found extending 15$^\circ$-20$^\circ$
from the plane \cite{Sivan}. A substantial H$\alpha$ emission is
observed from the Magellanic Stream \cite{Weiner}.

To first order, the H$\alpha$ may be modelled
as a layer parallel to the Galactic plane
with a half-thickness intensity of 1.2 Rayleigh (R).
The rms variation in this H$\alpha$ emission is about 0.6~R on degree
scales.
In the context of the present discussion 1~R
($\equiv {10^6 / 4 \pi}\,{\rm photons}\,{\rm cm}^{-2}\,{\rm s}^{-1}\,
{\rm sr}^{-1} = 2.42 \times 10^{-7}\,
\hbox{erg}\,\hbox{cm}^{-2}\,\hbox{sr}^{-1}\,\hbox{s}^{-1} $
at a wavelength $\lambda_{H\alpha} = 6563\,${\AA})
will give a brightness temperature of about 10$\,\mu$K at 45 GHz.

The situation is improving rapidly, however. The Wisconsin
H$\alpha$ Mapper (WHAM) survey \cite{Reynolds98} will produce a
Fabry-Perot map in H$\alpha$ of the northern sky with a $1^\circ$
resolution. Early results \cite{Haffner}
have revealed several very long ($\sim 30^\circ$--
$80^\circ$) filaments reaching high Galactic latitudes (up to $b \simeq
50^\circ$),
superimposed on the diffuse H$\alpha$ background,
without a clear correspondence to the other phases of the interstellar
medium, observed at 21cm or by radio continuum, far-IR or X-ray surveys.
High resolution ($\sim 1'$) imaging surveys of the H$\alpha$ emission in the
northern and in the southern emisphere, respectively, are underway by
\cite{Dennison} and \cite{Gaustad97}.

The four-year COBE DMR sky maps at different frequencies have been
utilized  \cite{Kogut} to isolate emission with antenna temperature
which varies proportional to frequency to the $-2.15$ power
($\propto \nu^{-2.15}$) in an attempt
to provide a large angular scale map of free-free emission at 53 GHz.
This low-signal to noise map is consistent with the H$\alpha$
large scale model with a free-free half height amplitude of $10 \pm
4\,\mu$K.
The rms free-free signal on a 7$^\circ$ scale was estimated to be
$\Delta T_{\rm ff} = 7 \pm 2 ~\mu$K.

Leitch et al. \cite{Leitch} looking near the NCP
found emission with an approximate $-2$ spectral index.
Their analysis indicates a free-free level some 5--10 times that found
for the NCP based upon an extrapolation in angular power spectra
from \cite{Kogut} or from the H$\alpha$ emission.
Moreover their estimates are above those deduced from the Tenerife
experiments.
Not surprisingly, the free-free emission is partially correlated
with dust as the H$\alpha$ emission has long been known to be.
However, there are things that are not understood well
and suprises may be lurking (see next section on dust).
Planck will clearly have a role in helping to understand the free-free
emission.

\subsubsection*{Dust emission}

The dust in our galaxy is heated by the interstellar
radiation field (ISRF), absorbing optical and UV photons and emitting the
energy in the far IR.
The balance between absorption and emission
depends on the intensity of the ISRF and the details of the dust, but
in general it results in an equilibrium temperature of 18--20$\,$K, in the
limit of large dust grains.
The emission spectrum from a greybody at this temperature peaks
at $\sim140\mu$m but extends out to microwave wavelengths.
Adequate removal of dust contamination from the microwave background signal
will allow both greater sky coverage and finer spatial resolution.

The new Berkeley-Durham dust map \cite{Schlegel}
is based on infra-red radiation from the dust,
observed by IRAS at $100\mu$m and by DIRBE at $100\mu$m and $240\mu$m.
The DIRBE $100\mu$m map was used to clean up and calibrate the IRAS Sky
Survey Atlas (ISSA) and $\sim5000$ point sources were removed using
positions from the 1.2Jy survey (Fisher et al. 1996).
A low-resolution temperature map was derived from the DIRBE $100\mu$m and
$240\mu$m data,
which is then applied to the high-resolution map.

Draine \& Lazarian \cite{DraineLaz98} have recently proposed that spinning
dust
grains can produce significant signal peaking in the few 10's of GHz.
Another mechanism of dust microwave emission is based on magneto-dipole
emissivity \cite{DraineLaz99}.
While there is no compelling evidence to support these hypotheses,
it is still a lesson to us that there could be suprising new
contributions or variation in Galactic emission awaiting us.
This indeed is a primary reason that Planck carries a complement
of frequencies.

\subsubsection*{Galactic point sources}

Planck measurements will detect the brightest
discrete Galactic radio sources,
bridging the gap between ground based observations, mostly confined to
frequencies $\leq 8\,$GHz, and far-IR data, mostly from IRAS and therefore
at $\nu \geq 3\times 10^3\,$GHz. As described below for some representative
classes of sources, Planck observations will provide crucial information
on their physical properties. Also these observations will be unique
in investigating the possibile existence of extremely compact thermal or
non thermal sources, self absorbed up to tens of GHz and therefore
unlikely to be found in existing surveys.

Optically thick synchrotron emission gives a strongly inverted spectrum
($S_\nu \propto \nu^{5/2}$), independent of the energy distribution
of electrons, producing a low frequency cutoff. For very compact, high
density regions, cutoff frequencies as high as several GHz are observed.
Likewise, compact and dense HII
regions may be optically thick, and have an inverted spectrum
($S_\nu \propto \nu^{2}$) up to relatively high frequencies.

{\it Supernova remnants} (SNR) constitute a particularly bright
and well studied class of Galactic radio sources. Nevertheless, existing
compilations
are believed to be very incomplete, especially with
respect to young, bright and older, extended remnants (see \cite{Weiler}
for a review).

The broad general category of SNRs actually
encompasses a variety of objects with widely different properties, which
have been divided into at least 4  subcategories. Their radio emission
is non-thermal, with spectral indices $\alpha$ ($S_\nu \propto
\nu^{-\alpha}$)
generally in the range 0.1--0.9. Planck is expected to detect a large
fraction of the 145 SNRs listed by Green (1984); more than half of them
will be resolved. Planck will be particularly effective in studying
the flat spectrum plerionic SNRs, which are possibly related to type II
SN and derive their energy from the rotational
energy losses of a central neutron star, rather than being powered by
the shock wave of the SN explosion, like the other classes of SNRs.
The Planck survey will greatly help in obtaining complete samples
of SNRs of the different classes, and in particular will allow to assess
the,
still debated, abundance of plerionic SNRs and to derive more reliable
estimates of their birthrates.  Moreover, the Planck mission will
provide accurate measurements of the continuum spectrum of the
brightest SNRs such as the Crab Nebula and Cassiopeia A, still poorly
known in the Planck frequency range. The Planck data
will be valuable to investigate the high frequency behaviour of
the synchrotron spectrum (and hence the energy distribution of
relativistic electrons within these SNRs) and the content of cold dust.

An outstanding mm source is {\it Sagittarius A}. This source,
located at the dynamical center
of the Galaxy and presumed to be associated with a supermassive
black hole ($M \sim 2.5\times 10^6\,\hbox{M}_\odot$ \cite{Ekart})
has been a puzzle for many
years (see e.g. \cite{Narayan}). It is brightest in the radio/mm
region. Its radio spectrum appears to have an ankle
at around $\nu\simeq 84\,$GHz, where it steepens from $S_\nu \propto
\nu^{0.2}$ to $S_\nu \propto \nu^{0.8}$, and is variable.
Planck covers this
particularly interesting region and will provide simultaneous measurements
at various frequencies.

An extensive survey of millimeter continuum emission from stars was carried
out by \cite{Altenhoff}. A review of the subject has been published by
\cite{Pallavicini}.
Massive radiatively driven winds from hot stars have mm radio spectra
dominated by optically thick free-free emission with $S_\nu \propto
\nu^{2/3}$. The two brightest sources of this class are $\eta$ Carinae
with a variable $1\,$mm flux of up to $16\,$Jy \cite{Cox}, making
it probably the brightest stellar source at mm wavelengths, and
MWC$\,$349 with a flux of $1\,$Jy at $3\,$mm.
Optically thick free-free emission
%($S_\nu \propto \nu^{0.65-0.80}$)
at mm wavelengths, with fluxes up to 1.5$\,$Jy, is also observed
from {\it symbiotic stars} \cite{Seaquist,Ivison},
a class of active binaries involving mass transfer from a red giant
to a compact, hot companion.
The transition to the optically thin regime occurs at about 100 GHz for
V1016 Cyg \cite{Seaquist} and, in general, should not exceed 1 THz,
i.e. is probably in the range covered by the Planck mission.
Hot stars with shells, an heterogeneous class of objects, frequently have
mm fluxes in the range 100-500$\,$mJy, reaching up to 1.6 Jy.
Their radio spectra usually have a turnover from optically thick to
optically thin free-free emission around 5--10 GHz.
The radio continuum spectra of {\it planetary nebulae} at GHz frequencies
are
generally consistent with optically thin
free-free emission from the circumstellar
material ionized by the hot central remnant. However
a few cases are known of strongly inverted spectra, indicating optically
thick emission up to 10 GHz (e.g. \cite{Knapp}).

\begin{figure}[t!] % fig 1
%\centerline{\epsfig{file=got.ps}}
\centerline{\epsfig{file=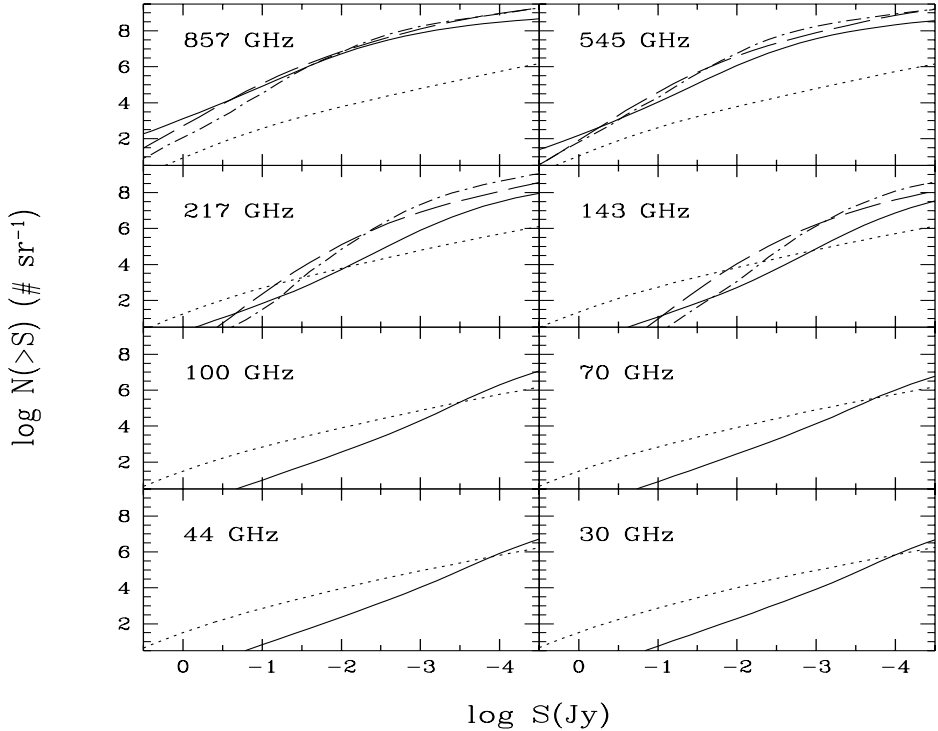,height=12.truecm,width=15truecm}}
\vspace*{-40pt}
\caption{Counts of extragalactic sources in Planck channels. The dotted
and the solid lines
shows the $\log N - \log S$ of radio sources and of dusty galaxies,
respectively, predicted by Toffolatti et al. (1998)
(see text).
The dashed and dot-dashed lines correspond, respectively,
to model E of Guiderdoni et al. (1998)
and to the kinematic model by Smail et al. (1997);
these models take into account only dusty galaxies. }
%\vspace*{10pt}
\label{fig1}
\end{figure}

\begin{figure}[t!] % fig 2
%\centerline{\epsfig{file=got.ps}}
\centerline{\epsfig{file=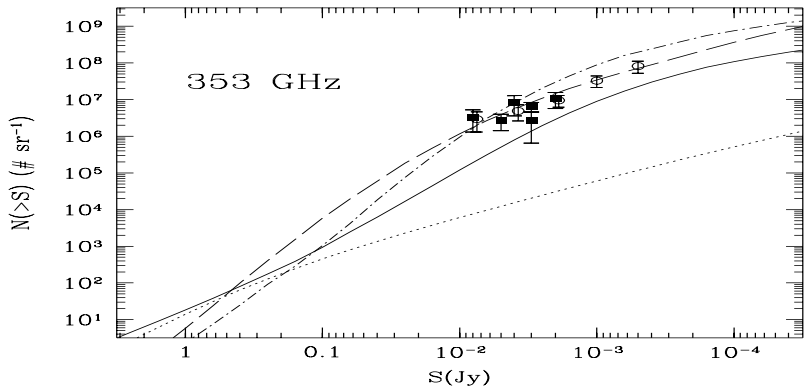,height=8.truecm,width=13truecm}}
\vspace*{-45pt}
\caption{A comparison of predictions by the same models as in Fig. 1 with
the $850\,\mu$m galaxy counts obtained by different groups (for
references see Smail et al. 1998)
with the Sub-millimeter Common User Bolometer Array
(SCUBA) on the 15-m James Clerk Maxwell Telescope (JCMT) at Mauna Kea. }
%\vspace*{10pt}
\label{fig2}
\end{figure}

\subsection*{Extragalactic sources}

The minimum of the spectral energy distribution for most classes of sources
occurs at a few mm wavelengths, i.e. close to the peak
frequency of the CMB. This is
due to the steep increase with frequency of the dust emission spectrum
in the mm/sub-mm region (typically $S_\nu \propto \nu^{3.5}$), which makes
the crossover between radio and dust emission components
only weakly dependent on their relative intensities; moreover,
 dust temperatures tend to be higher for distant high luminosity
sources, partially compensating for the effect of redshift. 

A consequence is that there is an abrupt change in the source
populations of bright sources observed in channels above and below
$\sim 1\,$mm: radio sources dominate at longer wavelengths, while
in the sub-mm region the Planck instruments will mostly see
dusty galaxies (see Figs. 1 and 2).

By simply looking for peaks more than $5\sigma$ above the mean, we can
recover
sources brighter than a few to several hundred mJy, the exact value
depending
on the angular resolution of each channel and on the model dependent
confusion noise (see e.g., \cite{Toffol}). It may be noted that
only a negligible fraction ($\simeq 3\times 10^{-7}$ in the case of gaussian
fluctuations) of true CMB anisotropies are misinterpreted as discrete
sources by this approach. 

Methods have been devised, allowing us
to recover point sources from Planck maps
to substantially fainter flux levels. Tegmark \& de Oliveira Costa
\cite{Tegmarkde} derived an optimal
filter in the Fourier domain, exploiting the point-like nature of sources,
as opposed to the diffuse CMB and Galactic emissions, to suppress the
latter components. They showed that this method improves the ability to
detect sources by factors between 2.5 and 18, depending on the channel.

Hobson et al. \cite{Hobson}
applied a maximum entropy method to analyse simulated
Planck observations. By introducing information about the power spectrum
of point sources in each frequency channel and the correlations between
frequencies, they were able to effectively recover point sources down to
60 mJy for the 44 and 353 GHz channels, chosen because they lie near the
centres of the frequency ranges of the LFI and HFI, respectively.

Ground based instruments, such as SCUBA on JCMT or the large mm/sub-mm
arrays planned for the next decade, or dedicated space missions, such as
FIRST, will go much deeper but over small areas. For example, the
SCUBA's field of view is $\sim 6$ arcmin$^2$, and the surveys carried
out so far have covered areas between 0.002 and 0.1 square degrees, so that
counts could be estimated only for sources fainter than about 10 mJy
(see Fig. 2).
The field of view of the Large Millimeter Array is expected to be
$\sim 50''$; about one month of observing time would be required to
survey 2 square degrees to a sensitivity of 0.9 mJy ($5\sigma$; see
\cite{Rowan}). The SPIRE photometer on FIRST, which
has three bands with nominal wavelengths of 250, 350, and 500$\,\mu$m,
has a $4'$ field of view; a survey of 4 square degrees to a $5\sigma$ limit
of 6.5 mJy would take about 1000 hours of observations \cite{Griffin}.

On the other hand, the full sky coverage of the Planck surveys makes it
possible to determine the source counts up to Jy levels, i.e. to extend
them by about two orders of magnitude in flux.
A broad range in fluxes is obviously essential to study the
epoch dependent luminosity function.

A further important foreground component of the microwave sky is produced
by the Sunyaev-Zeldovich effect in clusters of galaxies, whose flux changes
of sign at mm wavelengths, passing through 0 at 217 GHz.

%\begin{figure}[t!] % fig 3
\begin{figure} % fig 3
%\centerline{\epsfig{file=got.ps}}
\centerline{\epsfig{file=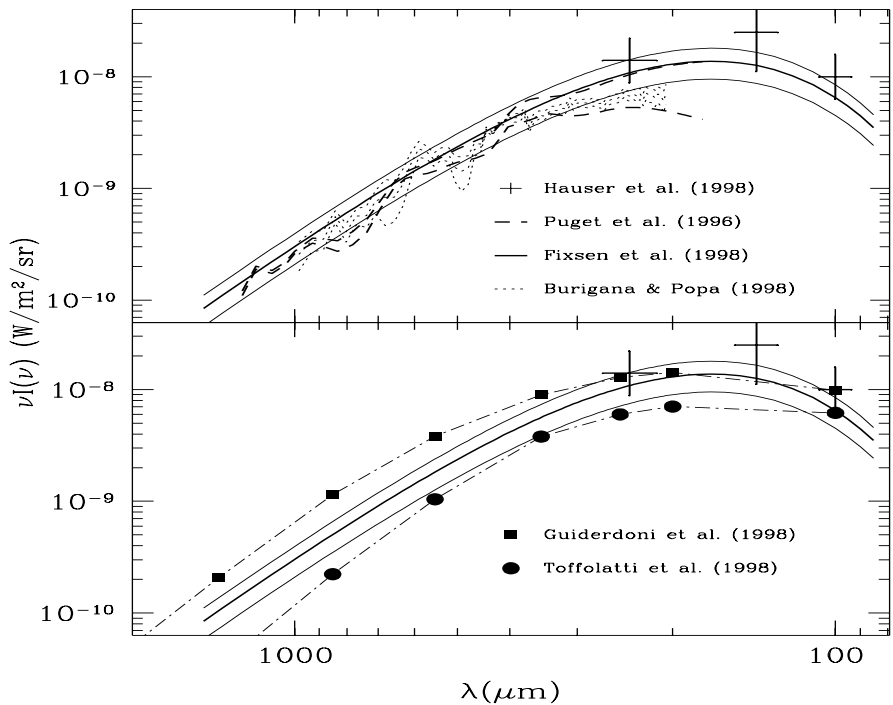,height=10.5truecm,width=15truecm}}
\caption{Recent determinations of the spectrum of the extragalactic far-IR
to mm background compared with model predictions.}
%\vspace*{10pt}
\label{fig3}
\end{figure}

\subsubsection*{Physics of Active Galactic Nuclei}

Planck covers the frequency range where the shape of the spectral energy
distribution of Active Galactic Nuclei is least known and where
important spectral features,
carrying essential information on physical conditions of sources,
show up.

Observations at
mm/sub-mm wavelengths often reveal the transition from optically thick
to optically thin radio emission in the most compact regions. The
self absorption frequency carries information on physical parameters.
A systematic survey in the 30--900 GHz range will, for example, allow
us to see if there are systematic differences in the synchrotron
turnover frequencies between e.g. BL Lacs and quasars, as would be
expected if BL Lac emission is angled closer to our line of sight so that
their turnovers are boosted to higher frequencies.
Correlations between turnover frequency and luminosity, which is
also boosted by relativistic beaming effects, would help confirm
current models.

Major high radio frequency flares have been observed in several
compact radio sources (e.g. PKS~0528$+$134, 3C~345, ...). There are
indications that the peak emission moves to lower frequencies as it
ages. Planck may catch the rise of the flare at the highest frequencies,
missed by ground based observations.

Establishing the peak of the synchrotron emission is crucial also to
understand if the emission at higher energies is to be attributed
to Compton scattering of the same synchrotron photons (synchrotron
self-Compton) or of seed photons external to the synchrotron
emitting region.

The spectral break frequency, $\nu_b$, at which the synchrotron
spectrum steepens due to electron energy losses, is related to the magnetic
field and to the ``synchrotron age'' $t_s$ (in Myr) by:
$\nu_b \simeq 96 (30\,\mu\hbox{G}/B)^{3}t_s^{-2}\,
\hbox{GHz}$.
The systematic multifrequency
study at the Planck frequencies will thus provide a
statistical estimate of the radiosource ages.

Excess far-IR/sub-mm emission, possibly due to dust, is often observed from
radio galaxies \cite{KnappP}. Planck data will allow us to assess
whether this is a general property of these sources;
this would have interesting implications
for the presence of interstellar matter in the host galaxies,
generally identified with giant ellipticals, which are
usually thought to be devoid of interstellar matter.

\subsubsection*{Flat-spectrum radio sources}

To date, the most detailed predictions of radio source counts at Planck
frequencies are those worked out by \cite{Toffol}. The dotted
line in Figs. 1 and 2 shows the predictions of a variant of their model,
whereby the spectra of ``flat''-spectrum sources keep an $\alpha =0$ slope
up to frequencies of 1 THz, steepening to $S_\nu \propto \nu^{-0.75}$
at higher frequencies.

Sokasian et al. \cite{Sokasian}
have produced skymaps of bright radio sources at
frequencies up to 300 GHz by means of detailed individual fits of the
spectra
of a large number of sources compiled from a number of catalogs, including
the all sky sample of sources with $S_{5\,{\rm GHz}} \geq 1\,$Jy
(K\"uhr et al. 1991). Their estimated number of sources
brighter than $S_{90\,{\rm GHz}} = 0.4\,$Jy is about a factor of two
below that predicted by \cite{Toffol}. A very similar result
was obtained by \cite{Holdaway} based on the observed distribution
of 8.4--$90\,$GHz spectral indices. On the other hand, these empirical
estimates yield, strictly
speaking, lower limits, since sources with inverted spectra may
be under-represented in the primary sample. Furthermore, in the presence of
substantial variability, estimates using mean fluxes underestimate
actual counts of bright sources \cite{DeZotti}. Anyway,
as mentioned by \cite{Sokasian}, it is very encouraging that so
different approaches yield results close to each other.

As shown by Fig. 1, the models by \cite{Toffol} predict that,
at least for fluxes brighter than 0.1 Jy, counts are dominated by radio
sources
for frequencies up to 217 GHz. These sources are therefore expected to have
a considerable impact on CMB anisotropy measurements.

\subsubsection*{Radio sources with strongly inverted spectra}

As already mentioned, classes of sources are known with
strongly inverted spectra, peaking at high frequencies,
that would be either missing from, or strongly
under-represented in low
frequency surveys.

GHz Peaked Spectrum radio sources (GPS) appear to have
a fairly flat distribution of peak frequencies extending out to 15 GHz in
the rest frame \cite{ODea}, suggesting the existence of
an hitherto unknown population of sources peaking at mm wavelengths
\cite{Crawford,Lasenby}.
De Zotti et al. \cite{DeZottietal} suggested
that from several tens to hundreds of GPS sources
could be detected by Planck.
Thus, although these rare sources will not be a threat for studies
of CMB anisotropies, we may expect that the Planck surveys
will provide crucial information about their
properties. GPS sources are important
because they may be the younger stages of radio source evolution
\cite{Fanti,Readhead} and may thus provide insight
into the genesis of radio sources; alternatively, they
may be sources which are kept very compact by unusual conditions
(high density and/or turbulence) in the interstellar medium of the
host galaxy \cite{vanBreu}.

High frequency free-free self absorption cutoffs may be present in AGN
spectra. Ionized gas in the nuclear region free-free absorbs radio photons
up to a frequency:
\begin{equation}
\nu_{\rm ff} \simeq 50 {g \over 5} {n_e \over 10^5\,{\rm cm}^{-3}}
\left({{\rm T}\over 10^4\,{\rm K}}\right)^{-3/4} l_{\rm pc}^{1/2}\ {\rm
GHz}\ .
\end{equation}
Free-free absorption cutoffs at frequencies $> 10\,$GHz may indeed be
expected, in the framework of the standard torus scenario for
type 1 and type 2 AGNs, for radio cores seen edge on, and may have been
observed in some cases \cite{Barvainis}. They
provide constraints on physical conditions in the parsec scale accretion
disk
or infall region for the nearest AGNs.

Planck may also allow us to study the nearest examples of
another very interesting class of radio sources,
powered by advection-dominated accretion flows
\cite{NarayanY,FabRee,DiMatteo}. These may correspond to the final stages
of accretion in large elliptical galaxies hosting a massive black hole.
Their radio emission is characterized by an inverted spectrum
with $S_\nu \propto \nu^{0.4}$ up to frequencies of 100--200 GHz,
followed by fast convergence.

\subsubsection*{Evolving dusty galaxies}

Shown in Figs. 1 and 2 are the counts predicted by the models of
\cite{Toffol}, \cite{Guiderdoni} (model E) and \cite{Smail}.
The former model falls somewhat short of the recent SCUBA counts at
$850\,\mu$m (Fig. 2) as well as of the far-IR to mm extragalactic background
intensity (Fig. 3) determined by \cite{Hauser,Fixsen}.
On the other hand, the model by \cite{Guiderdoni} overpredicts
the background intensity at $\lambda > 350\,\mu$m. The model by \cite{Smail}
yields a background intensity far in excess of the observational limits
and will not be considered further.

\subsection*{The Sunyaev-Zeldovich effect}

Another important astrophysical   foreground for CMB
observations are clusters of galaxies. As predicted
by \cite{su} the inverse Compton scattering of
 CMB photons against the hot and diffuse  electron gas
trapped in the potential well of cluster of galaxies, and
responsible for their    X-ray emission,  yields a systematic
shift of photons from the Rayleigh-Jeans to the Wien side of the
spectrum. This ``thermal" effect,  arising from the thermal
motions of the electrons,  is described by  the Kompaneets \cite{ko}
equation which,  in the non relativistic limit [for a discussion of
the relativistic Sunyaev \& Zel'dovich (hereafter SZ) effect see, in
this volume, Itoh, Rephaeli and Sunyaev],
leads to the familiar expression for the (spectral)
intensity change across a cluster:
\begin{equation}
\Delta I_\nu = 2{(k T_{CMB})^3 \over (hc)^2 }y g(x)
%\eqno(1)
\end{equation}
 where $T_{CMB}$ is the CMB temperature and
$x=h\nu/kT_{CMB}$. The spectral form of this ``thermal effect" is
described by the function
$g(x) = x^4\hbox{e}^x [x\cdot \coth(x/2) -4]/(\hbox{e}^x -1)^2$,  which is
negative
(positive) at values of $x$ smaller (larger) than
$x_0=3.83$, corresponding to a critical frequency  $\nu_0=217$ GHz.
The Comptonization parameter is
\begin{equation}
y = \int {kT  \over mc^2} n_e \sigma _T dl,
%\eqno(2)
\end{equation}
 where $n_e$ and $T $ are the electron density and
temperature, $\sigma _T$  is the Thomson cross section, and
the integral is over a line of sight  through the cluster.

The main emphasis so far has been on the measurement of the
SZ effect in individual clusters [see \cite{bi1} for a review and
Birkinshaw (this volume)]. On the other hand, because of its
 angular resolution and
sensitivity, Planck is expected to observe a substantial number of
clusters.
In fact,
with respect to the incident radiation field, the change of the
CMB intensity  across a cluster can be viewed as a net flux
emanating from the cluster, negative below the crossover
frequency  $\nu_0$  and positive above  it:
\begin{equation}
\overline{\Delta F}_\nu (\hat \gamma_\ell) =2 {(kT_{CMB})^3\over
(hc)^2}
y_o(\hat \gamma_\ell;M,z)\Xi (\hat \gamma_\ell; M,z)  \Biggl[\int
   dx g(x) E(x) /\int dxE(x) \Biggr]
%\eqno(3)
\end{equation}
In the
previous expression $M$ is the cluster mass, $\hat \gamma_\ell$ is the
line of
sight passing through the cluster center, $y_o$ is the
Comptonization parameter along it,
 $E(x)$ is the frequency response of the receivers and the geometrical
factor $\Xi$ is given by
 \begin{equation}
 \Xi (M,z) = \int d\Omega R_s(|\hat \gamma_l - \hat
\gamma|) \zeta(\hat\gamma,M,z)
%\eqno(4)
\end{equation}
 where $\zeta$ is the   intra-cluster (IC) gas profile and  $R_s$
is the angular response of the antenna.
%(hereafter assumed to be a gaussian with dispersion$\sigma_B$);

%hereafter assumed to be uniform inside the receiver pass-band.
Thus, the predicted number counts for SZ clusters are simply:
\begin{equation}
 N(>\overline {\Delta F_\nu}) = \int {dV \over dz} dz
\int_{\overline {M} (\overline {\Delta F_\nu},z)} dM N(M,z)
\,
%,\eqno(5)
\end{equation} where $N(M,z)$ is the cluster mass function and the lower
bound of the mass integral
  is determined by requiring that clusters of mass
$M>\overline M$ at redshift $z$ have  SZ fluxes (averaged
over the  receiver  pass-band and  measured by an antenna of
given angular resolution)   greater than ${\overline {\Delta
F_{\nu}}}$ (for details see, e.g., \cite{co97}).

\begin{figure}[t!] % fig 1v
\centerline{\epsfig{file=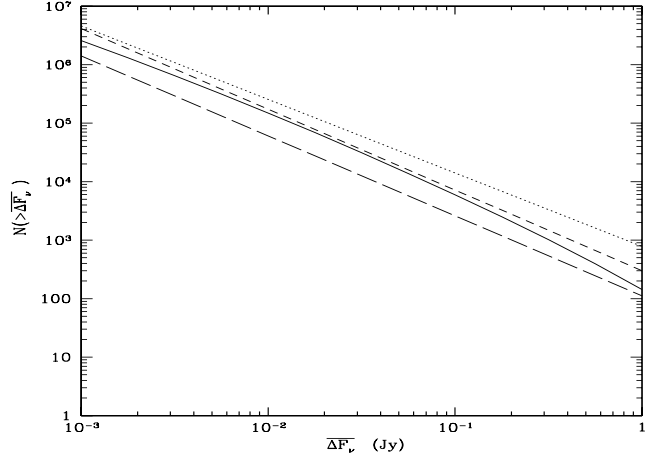,height=6.5truecm,width=3.5in}}
\vskip10pt
\caption{Comparison between the number counts in a flat universe calculated
according to Barbosa et al.
(dotted line), De Luca et al. (dashed line), Eke et al. (long-dashed line),
Colafrancesco et al. (continuous
line). These counts refer to the frequency of 400 GHz, except for those
by Eke et al., which refer to 545 GHz. Note that a SZ source is only roughly
half as intense at 545 GHz as it is at 400 GHz; this partially explains
why the Eke et al. curve is low. }
\label{fig1v}
\vspace*{10pt}
\end{figure}

Theoretical calculations of SZ cluster counts have been
performed using a Press \& Schecter \cite{pr574} formalism by
\cite{de95,ba96,eke96,co97}.
 In Fig.~\ref{fig1v} we show  a comparison of the
published SZ counts for a flat cosmological model with
Hubble constant of
$H_0= 50 \hbox{km}\,\hbox{s}^{-1}\,\hbox{Mpc}^{-1}$.
The differences in the SZ number counts are due to different assumptions and
to the large  uncertainties that we still have in the theoretical
modelling
of the effect. In short, the differences are due to the following points:
\begin{itemize}
 \item{}  {\it  density fluctuation power spectrum}: \cite{de95}
and \cite{ba96} use a simple power law power spectrum
[$P(k)= A k^n$]
with spectral index of $n=-1$ and $n=-1.85$, respectively; \cite{eke96}
and \cite{co97}  use scale-invariant initial
conditions
and take into account the transfer function appropriate for
Cold Dark Matter-like models [$P(k) = A k T_{CDM}^2(k)$].

\item{}  {\it Normalization of the density fluctuation
power spectrum}: The amplitude of the power
spectrum is normalized a posteriori in order to fit cluster
observations; \cite{de95} normalize this amplitude in order to recover the
cluster mass function of Bahcall \& Cen \cite{Bahcall}, while \cite{ba96}
normalize it to the cluster temperature function of \cite{he91};
\cite{co97} normalize $A$ to the X-ray luminosity
function of \cite{ko84} (see also \cite{co94}),
while \cite{eke96} derive the integrated cluster temperature function
from the original \cite{he91} catalogue and use this function to normalize
$A$.

\item{} {\it Mass-Temperature relation}: simple
scaling laws are   used to derive, under the hydrostatic
equilibrium hypothesis, the relation between the mass and the
temperature of a cluster (see, e.g., \cite{ba96}):
\begin{equation}
T(M,z)= T_{15}
\Omega_0^{1/3}h^{2/3}\Biggr[{\Delta_c(\Omega_0,z)\over
180}\Biggl]^{1/2}
\Biggr({M\over M_{15}}\Biggl)^{2/3}  (1+z)
%\eqno(6)
\end{equation}
 Here $\Delta_c$ is the non-linear  density contrast of a
cluster that collapsed at redshift $z$, while
$T_{15}$ is the temperature of a cluster of
$M_{15}=10^{15}h^{-1}M_\odot$ which collapses today.
De Luca et al. and Barbosa et al. \cite{de95,ba96}
renormalize $T_{15}$ to $6.8\,$keV, a value
suggested by the numerical simulation of \cite{ev90a};
Colafrancesco et al. \cite{co97}
normalize to $5.8\,$keV,  while
\cite{eke96} use for $T_{15}$ the value of  $7.5\,$keV, which is
  in good agreement with their N-body simulations. Normalizing the
amplitude of
$P(k)$ to the cluster temperature function minimizes the effect of different
choices of $T_{15}$.

\item{} {\it intra-cluster mass fraction}: the IC gas mass, $M_g$, is often
assumed
to be a fixed fraction of the total cluster mass: $f_g\equiv M_g/M$. De
Luca et al. and  Barbosa et al. \cite{de95,ba96}
made their calculations assuming a constant IC
gas mass fraction $f_g=0.2$,  while Eke et al. and Colafrancesco et al.
\cite{eke96,co97} use $f_g=0.1$.
\end{itemize}

The modelling of the IC gas is a critical issue to be discussed.
According to a numbers of
theoretical models (see e.g. \cite{ka91,ev91,ca98})
$f_g$ is not just a constant and can be written as follows:
\begin{equation}
 f(M,z) = f_0 \Biggr({M\over
M_{15}} \Biggl)^\eta  (1+z)^\xi.
%\eqno(7)
\end{equation}
where $f_0$ is a constant and $\eta$ is assumed to be independent of
$M$: this is
a good approximation if one considers only massive clusters (see
\cite{ca97}).
This simple parameterization  (originally proposed by \cite{ca93})
together with simple scaling laws provides in the case of a flat
universe (see, e.g., \cite{ma}):
\begin{equation}
L\propto T^{2+3\eta} (1+z)^{3(1-2\eta-2\xi)/2 }.
%\eqno(8)
\end{equation}
>From the observational point of view, the luminosity/temperature
relation,
$L\propto T^\alpha$, for X-ray clusters has been derived by several authors.
As initially discussed by \cite{fa94},
the intrinsic dispersion of these $L-T$ relation is
drastically reduced when corrections are made for the existence of
cooling
flows, thus providing a determination of the slope  $\alpha \sim 2.6$,
with a precision $\gsim 5\%$ \cite{al98,ma98,ar98}.
Using the cluster catalogues of \cite{al98} and \cite{ma98}, Reichart et al.
\cite{re98} derived the values
$\eta=0.22^{+0.08}_{-0.07}$ and
$\xi=0.35^{+0.35}_{-0.33}$. This $\xi$ value, appropriate for a flat
universe,
lowers to $\xi=0.14^{+0.35}_{-0.33}$ for $\Omega_0\rightarrow 0$.
Thus, cluster
formation  is not a self--similar process: the mass of the IC gas is not
a fixed
fraction of the cluster total mass. The values of $\xi$ derived by
\cite{re98} provide very little evolution of the $L-T$ relation up to
redshift
$\lsim 0.5$, as found by \cite{mu97} with the ASCA
data.

Eke et al. \cite{eke98} showed that the cluster temperature function, both
locally and at $z\sim 0.3$ \cite{he97}, is best fitted by a
low-density,
vacuum dominated, CDM model with $\Omega_0=0.3$. According to these
authors,
a flat universe is ruled out at the 98\% confidence level \cite{he97,eke98}
(see, however, \cite{sa98}, who find $\Omega_0=0.85\pm 0.2$).
\begin{figure}[t!] % fig 2v
\centerline{\epsfig{file=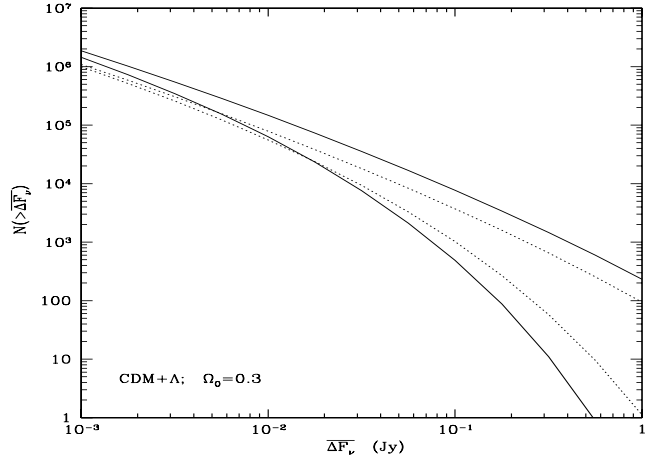,height=6.5truecm,width=3.5in}}
\vskip10pt
\caption{SZ cluster number counts for a low density,  vacuum-dominated
CDM model with $\Omega_0=0.3$. The
heavy lines represent the expected counts for the Planck/HFI 353 GHz
(continuos line) and Planck/LFI 100 GHz (dotted line) channels.
The light lines are the expected counts without
taking into account the receiver angular resolution.}
\label{fig2v}
%\vspace*{10pt}
\end{figure}
In Fig.~\ref{fig2v} we show the  SZ number counts expected for a
CDM+$\Lambda$ model
($\Omega_0=0.3$) with the angular resolution of  the Planck/HFI 353~GHz and
Planck/LFI 100 GHz channels, respectively. Resolution matters,
because the flux collected by an
antenna of angular resolution
$\sigma_{B}$, pointing at the cluster center, is only a fraction of the
 total one for those clusters whose extension is larger than
$\sigma_B$.  This effect, taken into account by \cite{co97},
tends to suppress cluster counts at high fluxes, as one looses
the less massive and more local clusters. These clusters may be
recovered by a
suitable smoothing of the original map, which will bring out large, coherent
regions with a higher signal to noise ratio. This effect is shown in
Fig.~\ref{fig2v} where we use $\eta=0.2$ and $\xi=0.3$: lowering
$\xi$ to 0.15 (the value found by \cite{re98}
for $\Omega_0=0$) increases the SZ number counts by only $10\%$.

The number of clusters that Planck will be able to observe critically
depends on
the efficiency in separating Planck observations into different physical
components (for a discussion on this point see \cite{ho98}
and references therein). In any case it should be possible to
select and measure a number of clusters, even without any component
separation.
The Planck detection of  SZ clusters
may provide an independent estimate of $\Omega_0$   (see, e.g., \cite{ba96};
see however \cite{vi98} for a different point of view),
complementing the more  precise one  arising from the measurement of the
angular
power spectrum of the primary  CMB anisotropy. The comparison of SZ and
X-ray
measurements will yield an  independent measure of the Hubble constant
\cite{ca77} (see \cite{bi1} for a recent review).
If the temperature structure of the cluster is simple, the
relation
between the SZ flux and the X-ray temperature of a cluster provides a
measure of
the IC gas mass, independently of how the IC gas is distributed. This is
due to the linear dependence of the SZ effect on the electron density.
Finally, the
correlation between the SZ signal and the X-ray luminosity will provide
an
interesting, complementary tool for studying the dependence of the IC
gas
properties on the cluster mass and the cosmological epoch \cite{co99}.

\begin{figure}[t!] % fig 6
%\centerline{\epsfig{file=got.ps}}
\centerline{\epsfig{file=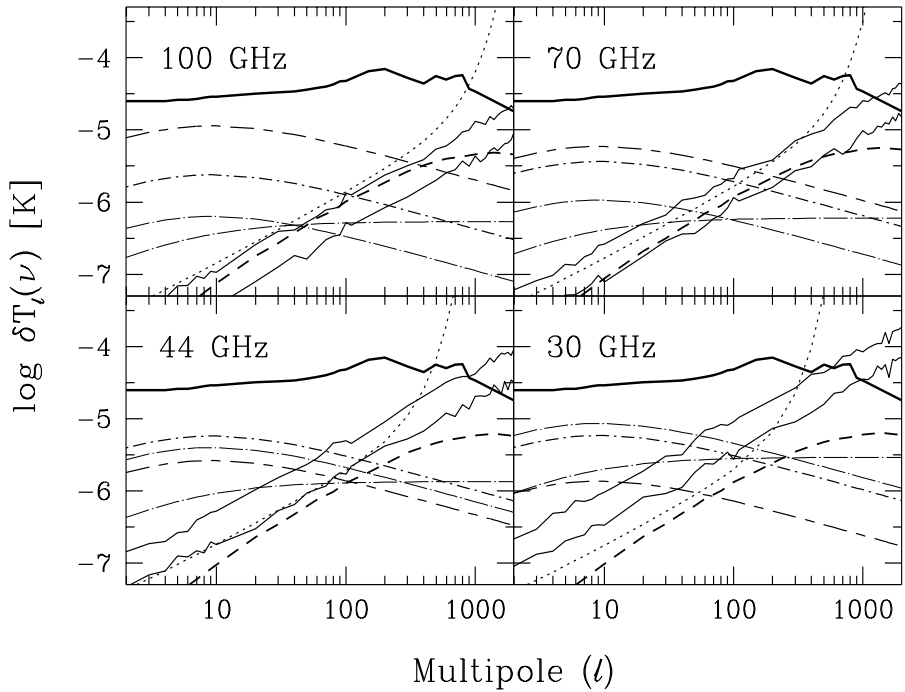,height=14truecm,width=13truecm}}
\vspace*{-80pt}
\caption{Angular power spectra of the components contributing to
fluctuations
at the LFI frequencies. Following Tegmargk \& Efstathiou (1996), we have
plotted, for each component, the quantity $\delta T_\ell(\nu) =
[\ell(2\ell+1)C_\ell(\nu)/4\pi]^{1/2}$.
The upper heavy solid curve shows the power spectrum of CMB
fluctuations predicted by the standard CDM model ($\Omega=1$,
$H_0= 50\,{\rm km}\,{\rm s}^{-1}\,{\rm Mpc}^{-1}$, $\Omega_b  =
0.05$).
Fluctuations in the Galactic emissions at $|b|>30^\circ$ are represented
by long$+$short dashes (dust), dots$+$short dashes  (free-free), and
dots$+$long dashes (synchrotron). For the power spectrum of the
synchrotron
emission we have plotted both the power spectrum derived by Tegmark
\& Efstathiou (1996)
and that observed in the Tenerife patch (Lasenby 1996);
the latter is significantly lower than the former in the range of scales
where it has actually been measured, but has a flatter slope so that it
may become relatively more important on small scales; it may be
noted however, that the power spectrum must fall off on scales smaller
than the coherence length of the emitting blobs. The roughly diagonal
solid lines show
the contributions of extragalactic sources, neglecting the effect of
clustering and assuming that sources brighter than 1 Jy (upper line)
or 0.1 Jy can be identified and
removed. The heavy dashed line peaking at large $\ell$ shows the power
spectrum of anisotropies due to the Sunyaev-Zeldovich effect computed
by Atrio-Barandela \& M\"ucket (1998)
adopting a lower limit of $10^{14}\,{\rm M}_\odot$ for
cluster masses, a present ratio $r_{\rm virial}/r_{\rm core}= 10$, and
$\epsilon = 0$. The dotted
line shows the unsmoothed noise contribution of the instrument. }
%\vspace*{10pt}
\label{fig6}
\end{figure}

\begin{figure}[t!] % fig 7
%\centerline{\epsfig{file=got.ps}}
\centerline{\epsfig{file=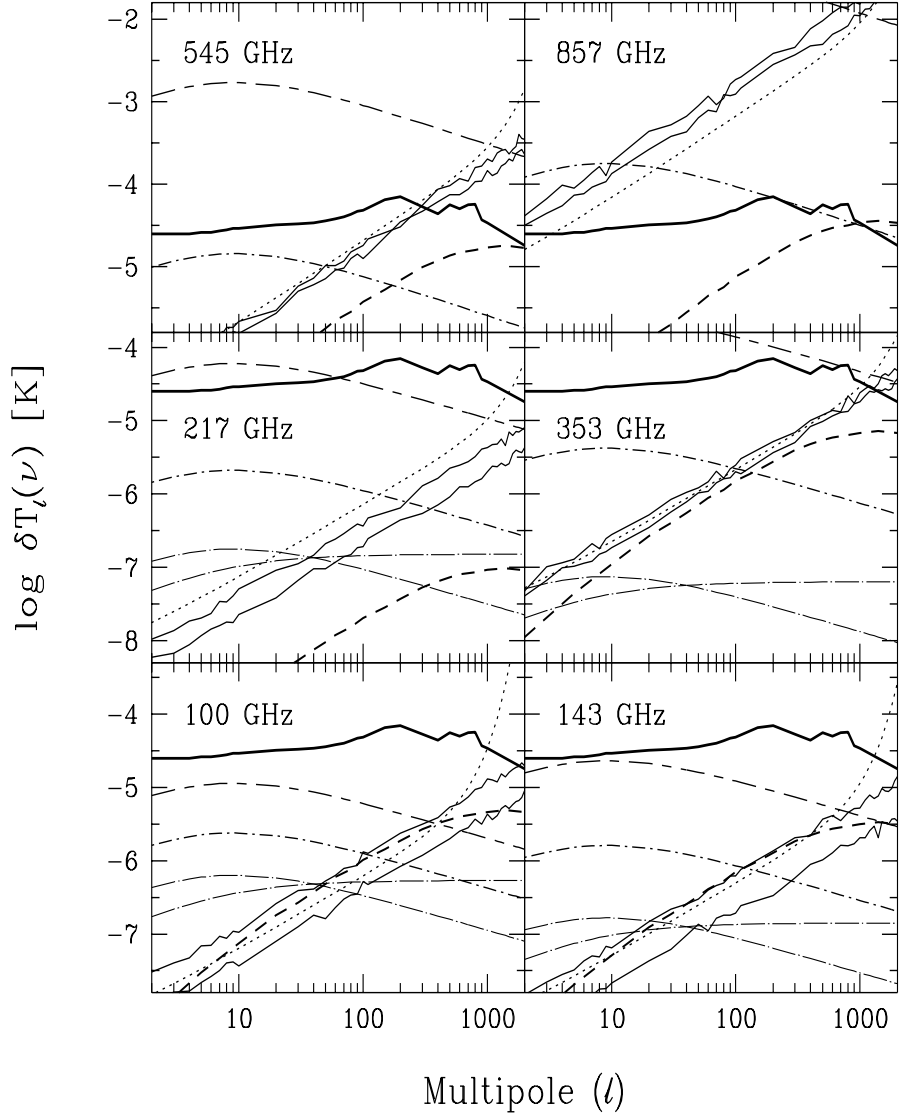,height=16truecm,width=13truecm}}
\vspace*{10pt}
\caption{Same as in Fig.~\ref{fig6}, but for HFI frequencies.
The lines have the same meaning as in Fig.~\ref{fig6}.
Note that the Sunyaev-Zeldovich effect vanishes at 217 GHz. However,
integrating over the bandwidth of the Planck/HFI channel centered at that
frequency, a small but non vanishing signal is found.}
%\vspace*{10pt}
\label{fig7}
\end{figure}

\section*{Power spectra of foreground fluctuations}

The angular power spectrum of the Galactic synchrotron emission
can be determined from large scale radio maps, particularly from those at
408 and 1420, after removing the baseline stripes which contain power
on angular scales of a few to ten degrees \cite{Davies}.
Lasenby \cite{Lasenby} has used the 408 and 1420 MHz surveys to estimate
the spatial power spectrum of the high latitude region surveyed
in the Tenerife experiments and found
an angular power spectrum slightly flatter than $\ell^{-2}$
for $\ell \leq 300$. A steeper mean power spectrum ($\propto \ell^{-3}$),
i.e. with the same dependence on $\ell$ as the dust and free-free
emissions, was estimated by \cite{TegmarkEf}.

Kogut et al. \cite{Kogut} have compared the COBE DMR maps
with the DIRBE maps and found a correlation between the free-free and
dust emission. On 7$^\circ$ angular scales
they concluded that the spatial power law index is $-3.0$,
a well-determined value for dust and HI on degree scales.
This was confirmed to some extent by \cite{Costa}
for the north celestial polar cap in a correlation of Saskatoon data
with the IRAS \& DIRBE maps.

The H$\alpha$ images of the North Celestial Pole (NCP)
area made by \cite{Gaustad}  have been
analyzed by \cite{Veeraraghavan} to provide an estimate
of the spatial power spectrum on scales of 10$^\prime$ to a few degrees.
The slope they find, $C_{\rm ff} \propto \ell^{-2.27\pm 0.07}$,
is significantly flatter than that inferred
by \cite{Kogut} from COBE data; however the rms amplitude  is
considerably lower, $\simeq 0.25{\rm cosec}(|b|)\,\mu{\rm K}$ on $10'$
scales
at 53 GHz assuming a gas temperature $\sim 10^4\,$K,
resulting in a lower signal at all scales of interest.

\begin{figure}[t!] % fig 8
%\centerline{\epsfig{file=got.ps}}
\centerline{\epsfig{file=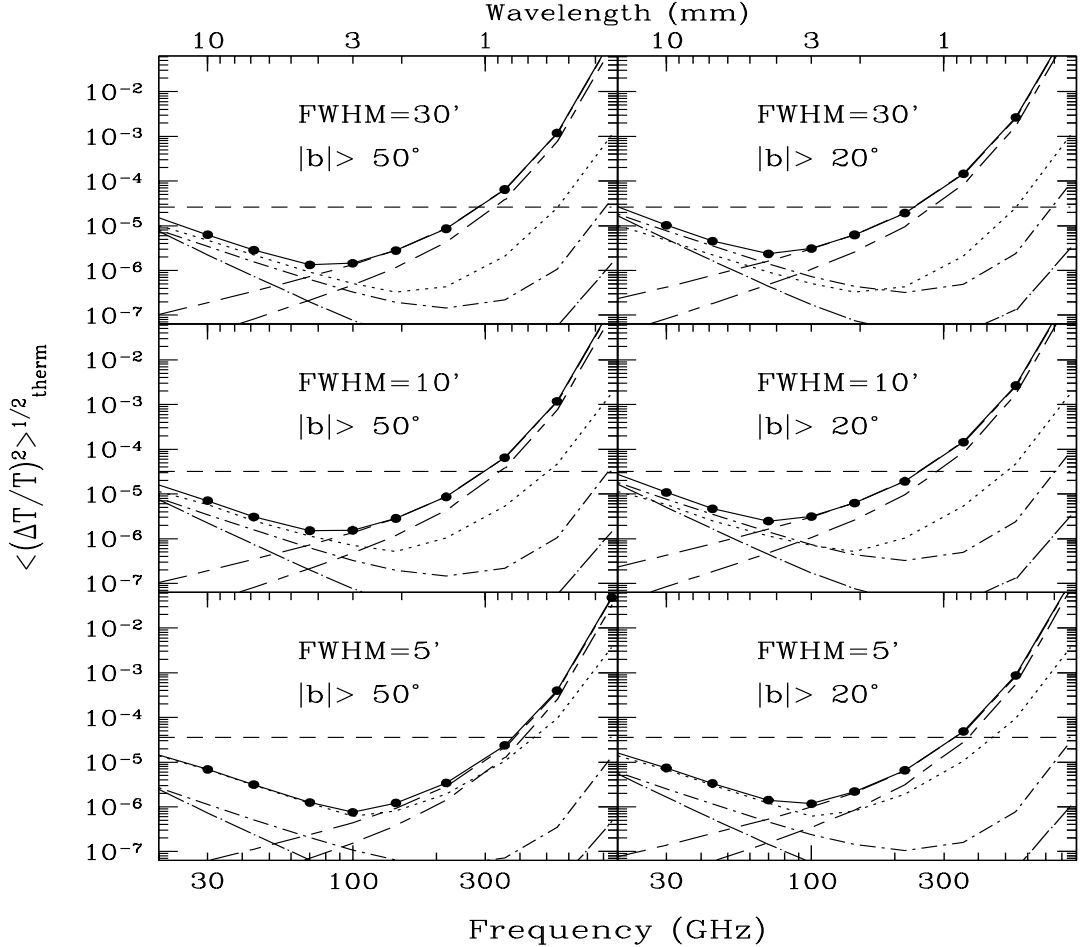,height=13truecm,width=16truecm}}
\caption{Temperature fluctuations as a function of frequency for three
angular scales and two cuts in galactic latitude. The horizontal dashed line
shows the expected level of primordial CMB anisotropies according to the
standard CDM model. The average contributions from Galactic free-free,
synchrotron and dust emissions are represented by dots$+$short dashes,
dots$+$long dashes, and long$+$short dashes, respectively. The lower
long/short dashed curve shows a lower limit to Galactic dust emission
fluctuations. The dotted curve gives the fluctuation level yielded by
discrete extragalactic sources fainter than 100 mJy, computed after
the Toffolatti et al. (1998) model (see text). The solid line is the
quadratic sum of contributions from all foreground components; the filled
circles on it mark the Planck frequencies. }
%\vspace*{10pt}
\label{fig8}
\end{figure}

\begin{figure}[t!] % fig 9
%\centerline{\epsfig{file=got.ps}}
\centerline{\epsfig{file=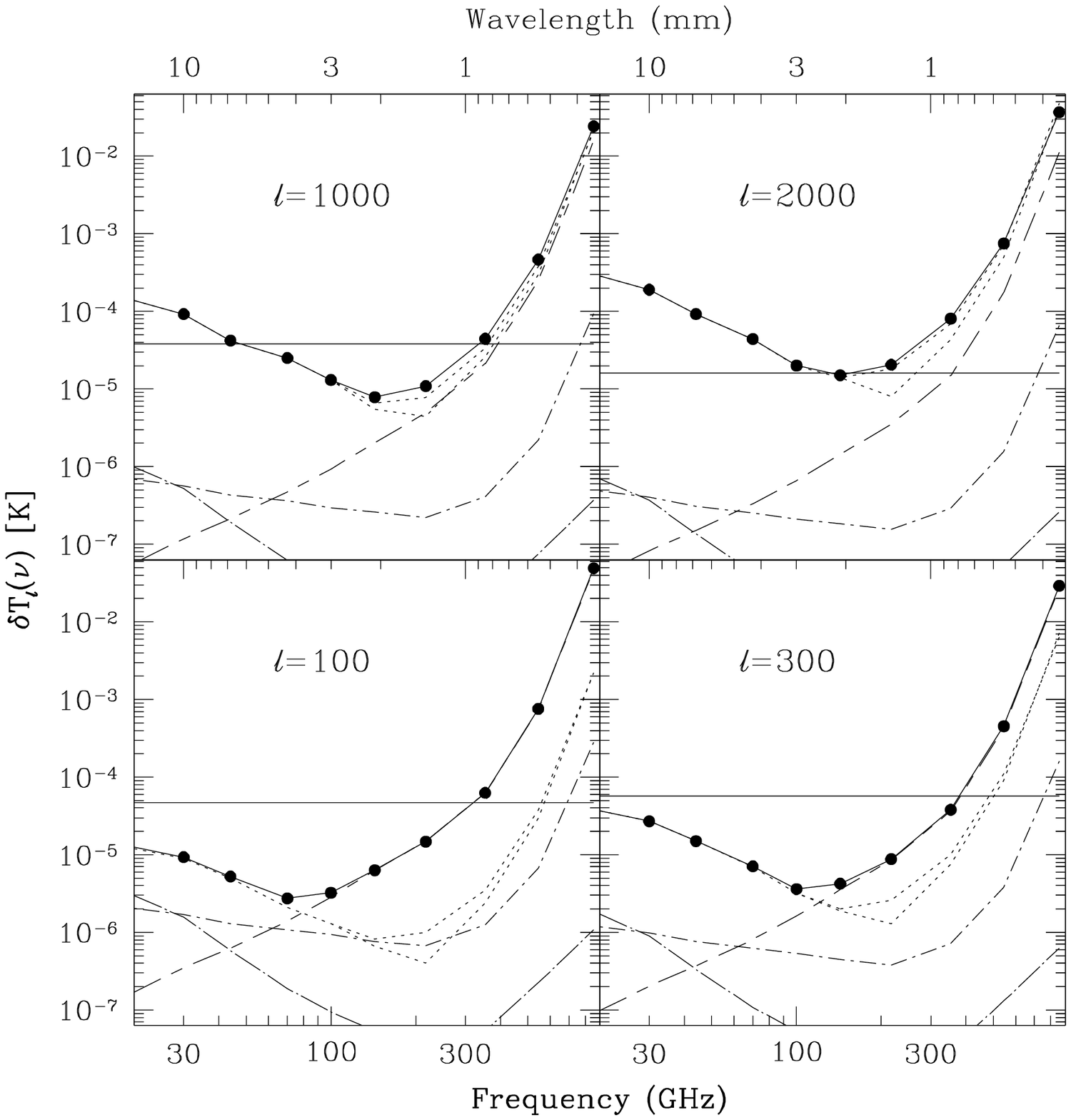,height=12truecm,width=15truecm}}
\vspace*{-10pt}
\caption{Frequency dependence of CMB brightness fluctuations
$\delta T_\ell(\nu) = [\ell(2\ell+1)C_\ell(\nu)/4\pi]^{1/2}$ for different
multipoles. The dots$+$long dashes, dots$+$short dashes, and long$+$short
dashes correpond to the mean contributions from Galactic foregrounds
(synchrotron, free-free and interstellar dust, respectively), at
$|b|>50^\circ$. The dotted lines show the contribution of extragalactic
sources (assuming that those brighter than 1 Jy are removed):
the lower one corresponds to the Toffolatti et al. (1998)
model mentioned in the text; the upper one was obtained summing in
quadrature the results of model E by Guiderdoni et al. (1998) for
evolving dusty galaxies with the results by Toffolatti et al. (1998)
for radio sources. The heavy solid curve is the quadratic sum of all
foreground contributions; the filled circles on it identify Planck
frequencies. The horizontal line shows the fluctuation level predicted by
the standard CDM model.}
%\vspace*{10pt}
\label{fig9}
\end{figure}

The global power spectrum of dust emission fluctuations was determined by
\cite{Gautier}, based on IRAS $100\,\mu$m data, to be $C_{\rm dust}
\propto \ell^{-3}$ in the $8^\circ$--$4'$ range.
Wright \cite{Wright}, from an analysis of COBE/DIRBE data with two methods,
also found, at high galactic latitude, $C_{\rm dust} \propto \ell^{-3}$ for
$2 < \ell < 300$. The Schlegel et al. \cite{Schlegel}
analysis of their combined DIRBE
and IRAS dust maps suggests a shallower slope, $C_{\rm dust} \propto
\ell^{-2.5}$, with variations from place to place.

For $\ell < 300$, corresponding to angular scales $> 30'$
[$\ell \simeq 180^\circ/\theta({\rm deg})$], diffuse
Galactic emissions dominate foreground fluctuations even at high Galactic
latitudes. These are minimum at $\nu \simeq 70\,$GHz \cite{Kogut}.
For larger values of $\ell$ the dominant contribution is from extragalactic
sources; their minimum contribution to the
anisotropy signal occurs around 150-200$\,$GHz.

A Poisson distribution of extragalactic point sources produces a simple
white-noise power spectrum, with the same power in all multipoles.
Correspondingly, the point source fluctuations become increasingly important
with decreasing angular scale (i.e., with increasing $\ell$). Thus
the minimum in the global power spectrum of foreground fluctuations
moves, at high galactic latitudes, from about 70 GHz for $\ell < 300$,
where Galactic emissions dominate, to
150-200$\,$GHz at higher values of $\ell$ (the exact value of the
frequency of the minimum depends on the detailed spectral and evolutionary
behaviour of sources). For the same reason,
the frequency of minimum foreground fluctuations
decreases with decreasing Galactic latitude.

Figures 6 and 7 show the expected power spectra of the main foregrounds
components in the LFI and HFI channels respectively.
It is clear from Fig.~7 that the source removal is much more
effective in reducing the fluctuation level at the lowest than at the
highest
frequencies. This is due to the fact that at $\lambda \gsim 1\,$mm,
fluctuations are dominated by the brightest sources below the detection
limit,
while at shorter wavelengths the dominant population are evolving dusty
galaxies whose counts are so steep (see Fig. 1) that a major contribution
to fluctuations comes from much fainter fluxes.

Figure 8
illustrates the frequency dependence of rms fluctuations in a beam of
FWHM $\theta$ for several values of $\theta$, while Fig. 9 shows the
frequency dependence of the power spectrum for four values of $\ell$.

\begin{figure}[t!] % fig 10
%\centerline{\epsfig{file=got.ps}}
\vspace*{-30pt}
\centerline{\epsfig{file=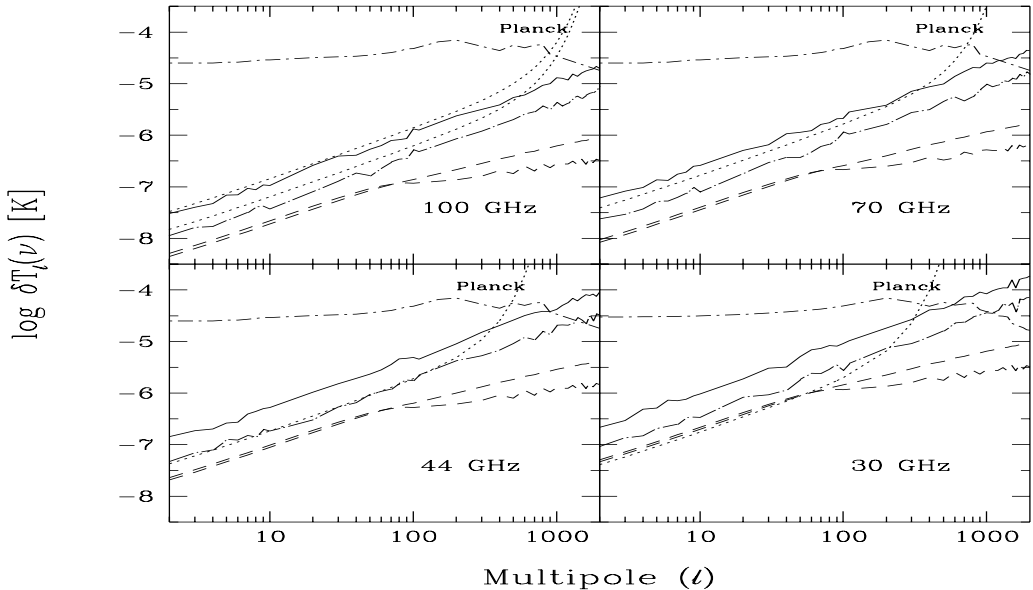,height=9truecm,width=15truecm}}
\caption{Comparison of the Poisson component of the power spectra of
radio sources with the component due to clustering. The upper dot/dashed
curve
shows the primordial CMB power spectrum according to the
standard CDM model. The roughly diagonal solid and dot/dashed lines
represent the Poisson component for flux cuts of 1 Jy and 0.1 Jy,
respectively.
The dashed lines show the component due to clustering estimated using
the angular correlation functions derived by Loan et al. (1997; lower line)
and
by Magliocchetti et al. (1998).
The dotted lines show the instrumental unsmoothed noise contributions; at
100 GHz there are two lines referring to the LFI (lower) and to the HFI.
Only
the LFI operates at the other frequencies. }
%\vspace*{10pt}
\label{fig10}
\end{figure}

\subsection*{The effect  of clustering}

Toffolatti et al. \cite{Toffol}
found that clustering contributions to fluctuations
due to extragalactic sources is generally small in comparison with the
Poisson
contribution. However, the latter contribution, at least in the case of
radio sources, comes mostly from the brightest sources below the detection
limit, while the clustering term is dominated by fainter sources. Therefore,
an efficient subtraction of sources decreases the Poisson term much more
effectively than the clustering term, which therefore becomes relatively
more and more important.

In the case of a power law angular correlation function ($w(\theta) \propto
\theta^{1-\gamma}$), the power spectrum of intensity fluctuations is
(eq. (58.13) of \cite{Peebles}):
\begin{equation}
C_\ell \propto \ell^{\gamma -3}\ .
\end{equation}
If this behaviour extends to large enough angular scales,
i.e. to small enough values of $\ell$, the clustering
signal will ultimately become larger than the Poisson anisotropy. On the
other hand, for large values of $\theta$, $w(\theta)$ is expected to drop
below the above power law approximation, and $C_\ell$ will
correspondingly break down.

The preliminary estimates (Toffolatti et al., in
preparation) reported in Fig. 10 are based on the
correlation functions of radio sources
derived by \cite{Loan} and by \cite{Magliocchetti}.
The different slopes at large values of $\ell$ reflect
different values of $\gamma$: the upper curve correspond to the larger
value ($\gamma = 2.5$) obtained by \cite{Magliocchetti}.

Scott \& White \cite{Scott}
have recently shown that if dusty galaxies cluster like
the $z \sim 3$ Lyman break galaxies, at frequencies $\geq 217\,$GHz
the anisotropies due to clustering may exceed the Poisson ones on all scales
accessible to Planck; in the 353$\,$GHz ($850\,\mu$m) channel the clustering
signal may exceed the primordial CMB anisotropies on scales smaller than
about $30'$. It should be noted, however, that current models
\cite{Toffol,Guiderdoni} strongly suggest a broad redshift
distribution of sources contributing to the autocorrelation function of
the intensity fluctuations, implying a strong dilution of the
clustering signal. A further substantial overestimate of the effect of
clustering may follow from the extrapolation to degree scales, with
constant slope, of the angular correlation function determined on
scales of up to a few arcmin.
Indeed, a clustering signal much lower than estimated
by \cite{Scott} was found by \cite{Toffol} who modelled
the full evolutionary history of galaxies.

\section*{Polarization}

The polarized foreground spectra have not been studied as extensively as the
intensity spectra. For a power law electron energy spectrum $dN/dE=N_0
E^{-p}$,
the polarization level, $\Pi$, yielded by a
uniform magnetic field is \cite{LeRoux,Ginzburg}:
\begin{equation}
\Pi = {3(p +1) \over 3p +7} \ .
\end{equation}
For a typical high frequency value of $p \sim 3$, $\Pi\sim 75\%$.
Non-uniformities of the magnetic fields and differential
Faraday rotation decrease the
polarization level. However, the Faraday rotation optical depth is
proportional to $\nu^{-2}$ so that Faraday depolarization is negligible
at the high frequencies relevant for Planck.

{\it Free-free emission}
is not polarized. However, Thompson scattering by
electrons in the HII regions where it is produced, may polarize it
tangentially to the edges of the electron cloud \cite{Keating}. The
polarization level is expected to be small, with an upper limit of
approximately 10\% for an optically thick cloud.

The intrinsic polarization of {\it Galactic dust emission} is estimated to
be
$\sim 30\%$ \cite{Hildebrand}. The observed polarization degree
is smaller by a factor $\Phi =RF\cos^{2}\gamma$ \cite{Lee}, where
$R$ and $F$ are the reduction factors due to misalignments of grain axes
with the magnetic field and to the different orientations of polarization
vectors of different components along any line of sight, while the
$\cos^2\gamma$ accounts for the projection of the direction of polarization
on the plane of the sky. The level of polarized emission from Galactic dust
at high Galactic latitudes has been estimated by \cite{Prunet}. To ease
the comparison with the CMB polarization, they considered two linear
combinations of the Stoke's parameters in Fourier space: the E-mode,
dominated
by scalar perturbations, and the B-mode, produced by tensor perturbations.
For Galactic dust polarized emission, the power spectra of the two
modes can be approximated by \cite{Prunet}:
\begin{equation}
C_E(\ell) = 8.9\times 10^{-4} \ell^{-1.3}\ (\mu{\rm K})^2
\end{equation}
\begin{equation}
C_B(\ell) = 1.\times 10^{-3} \ell^{-1.4}\ (\mu{\rm K})^2
\end{equation}
They concluded that, at frequencies around 100 GHz, the power spectrum of
this polarized component is below that of the scalar induced CMB
polarization
at $\ell \gsim 200$, but is above the tensor induced CMB polarization
expected for a flat, tilted cold dark matter model with spectral index of
scalar perturbations $n_s=0.9$.

Bouchet et al. \cite{Bouchet}
have analized the possibility of extracting the power
spectrum of CMB polarization fluctuations in the presence of polarized
Galactic foregrounds using a multifrequency Wiener filtering of the data.
They concluded that the power spectrum of E-mode polarization of the CMB
can be extracted from Planck data
with fractional errors $\lsim 10$--30\% for
$50 \lsim \ell \lsim 1000$. The B-mode CMB polarization, whose detection
would unambiguously establish the presence of tensor perturbations
(primordial
gravitational waves), can be detected by Planck with signal-to-noise
$\simeq 2$--4 for $20 \leq \ell \leq 100$ by averaging over a 20\%
logarithmic range in multipoles.

Polarization of {\it extragalactic sources}, not considered by
\cite{Bouchet},
can be an important issue as well. Flat spectrum radio
sources are typically 4-7\% polarized at cm and mm wavelengths
\cite{Nartallo,Aller}. For random orientations of the magnetic
fields of sources along the field of view, the rms polarization
fluctuations are approximately equal to intensity fluctuations times
the mean polarization degree (De Zotti et al., in preparation).

\section*{Conclusions}

As shown by Figs. 6 and 7, the
experimental accuracy of the Planck Surveyor mission is effectively
limited by astrophysical foregrounds. At high Galactic latitudes,
however, fluctuations due to the Galactic dust, free-free and
synchrotron emission, which are the main foreground components for
angular scales $\gsim 30'$, are expected to be well below the primordial
CMB anisotropies in several Planck channels. On these angular scales,
i.e. for multipoles $\ell \lsim 300$, foreground fluctuations are minimum
at $\simeq 70\,$GHz (see Figs. 8 and 9).

On smaller scales, the most troublesome foreground component are
extragalactic
sources. For wavelengths $\gsim 1\,$mm, the dominant population is made
of flat-spectrum and possibly of some inverted spectrum radio sources;
estimates using very different methods indicate that their counts can be
evaluated, at the fluxes relevant to estimate fluctuations in Planck
channels, to better than a factor of 2 at frequencies up to at least 100
GHz.
The uncertainty is larger for
evolving dusty galaxies which are expected to dominate the counts at
shorter wavelengths, but important constraints are set by SCUBA counts and
by determinations of the extragalactic mm/sub-mm background spectrum.
Point source fluctuations are larger than CMB anisotropies on very small
angular scales ($\ell > 2000$, see Fig. 9). In the frequency region 100--200
GHz, however, their amplitude is significantly below the CMB signals for
all scales accessible to Planck instruments.

Fluctuations due to clustering are generally small in comparison
with Poisson fluctuations, at least in the case of radio sources (Fig. 10).
Recent results indicating, in the case of evolving dusty galaxies, a
clustering signal exceeding Poisson fluctuations, are likely to be
rather extreme upper limits.

CMB polarization measurements are more challenging and the contamination by
polarized foreground emissions has not been studied in depth. However,
preliminary results suggest that the extraction of the CMB polarization
is possible, with a signal to noise ratio $>3$, for $50 \lsim
\ell \lsim 1000$.

On the other hand, the study of foregrounds is of great interest per se.
Planck will carry out calibrated all-sky surveys at 9 frequencies
between 30 and 860 GHz, covering an essentially unexplored spectral region.
>From several hundred to many thousands of sources can be detected
as $\geq 5\sigma$ peaks at each frequency. More refined analyses, exploiting
the different spectral properties and the different angular scales of
sources
in comparison with CMB fluctuations, may increase the number of detectable
source by substantial factors.

Planck will provide the best calibrated all sky
maps of the diffuse Galactic emissions (synchrotron, free-free and dust),
of the Sunyaev-Zeldovich effect in clusters of galaxies and of
extragalactic sources. It will extend the counts of the latter
by about two orders of magnitude in flux.
The information provided by Planck surveys will be unique:
large area radio continuum surveys above 30 GHz are not feasible from the
ground since the beam area of a given telescope scales as $\nu^{-2}$.

Planck measurements will also determine the
spectral energy distribution of bright sources over a factor $\simeq 30$ in
frequency, covering a region where spectral features carrying essential
information on physical conditions of sources show up
(self-absorption turnovers
of very compact components, high frequency flares, breaks due to energy
losses of relativistic electrons, ...).
While lower frequency surveys provide much more detailed information
relevant to define {\it phenomenological} evolution properties, surveys
at mm wavelengths are unique to provide information on the {\it physical}
properties.

Planck will also provide the first complete samples
of the extremely interesting classes of extragalactic radio sources
characterized by inverted spectra (i.e. flux density increasing with
frequency), which are very difficult to detect, and therefore
are either missing from, or strongly
underepresented in low
frequency surveys and may be very difficult to distinguish
spectrally from fluctuations in the CBR \cite{Crawford}.

Strongly inverted spectra up to tens of GHz can be produced
in very compact, high electron density regions, by
optically thick synchrotron emission or by free-free absorption.
Examples are known also among galactic sources.

In conclusion, extragalactic sources will not be a threat to Planck's
cosmological investigations. At the same time, Planck will provide
extremely interesting data for astrophysical studies.

\end{document}